\documentclass[12pt]{article}
\usepackage{graphicx}
\usepackage{booktabs,multirow} % Top and bottom rules for table
\usepackage{authblk}
\usepackage{natbib} % Required to change bibliography style to APA
\usepackage{amsthm,amsfonts}
\usepackage{epstopdf}
\usepackage{caption}
\usepackage{subcaption,amsmath} 
\usepackage{color}
\usepackage{bm}
\newtheorem{theorem}{Result}
\newcommand{\Mod}[1]{\ (\mathrm{mod}\ #1)}

\newcommand{\reva}[1]{{\color{black} #1}} %%  blue
\newcommand{\revb}[1]{{\color{black} #1}} % red
\newcommand{\blind}{0}

\begin{document}
\def\spacingset#1{\renewcommand{\baselinestretch}%
{#1}\small\normalsize} \spacingset{1}

%%%%%%%%%%%%%%%%%%%%%%%%%%%%%%%%%%%%%%%%%%%%%%%%%%%%%%%%%%%%%%%%%%%%%%%%%%%%%%
  %\thanks{
%The authors gratefully acknowledge}\hspace{.2cm}\\
\if0\blind
{
  \title{\bf Constructing Large Orthogonal Minimally Aliased Response Surface Designs by Concatenating Two Definitive Screening Designs}
\author[1,*]{Alan R. Vazquez}
\author[2, 3]{Peter Goos}
\author[2]{Eric D. Schoen}
\affil[1]{School of Engineering and Sciences, Tecnologico de Monterrey, Mexico}
\affil[2]{Department of Biosystems, KU Leuven, Belgium}
\affil[3]{Department of Engineering Management, University of Antwerp, Belgium}
\affil[*]{Corresponding author. Email: {alanrvazquez@tec.mx};}
\affil[ ]{Contributing authors: {peter.goos@kuleuven.be}, {eric.schoen@kuleuven.be}}
  \maketitle
} \fi

\if1\blind
{
  \bigskip
  \bigskip
  \bigskip
  \begin{center}
    {\LARGE\bf \bf Constructing large Orthogonal Minimally Aliased Response Surface designs by concatenating two Definitive Screening Designs}
\end{center}
  \medskip
} \fi

% ---------------------------------------------------------------------------------
% 						Abstract
% ---------------------------------------------------------------------------------
\begin{abstract}
\noindent Orthogonal minimally aliased response surface (OMARS) designs permit the study of quantitative factors at three levels using an economical number of runs. In these designs, the linear effects of the factors are neither aliased with each other nor with the quadratic effects and the two-factor interactions. Complete catalogs of OMARS designs with up to five factors have been obtained using an enumeration algorithm. However, the algorithm is computationally demanding for designs with many factors and runs. To overcome this issue, we propose a construction method for large OMARS designs that concatenates two definitive screening designs and improves the statistical features of its parent designs. The concatenation employs an algorithm that minimizes the aliasing among the second-order effects using foldover techniques and column permutations for one of the parent designs. We study the properties of the new OMARS designs and compare them with alternative designs in the literature. 

\smallskip


\end{abstract}

\pagebreak
% ---------------------------------------------------------------------------------
% 						Introduction
% ---------------------------------------------------------------------------------
\spacingset{1.45} 

\section{Introduction}
\label{sec:intro}

% Screening with available designs (optional)

Many industrial experiments are designed to study the effect of multiple quantitative factors on one or more responses. For example, \cite{Maestronietal2018} designed a screening experiment to study the effect of seven factors on the robustness of a method to extract pesticides from potato. Table~\ref{tab:factors} shows the factors. They include an agitation time and two shaking times of specific compounds at different stages of the extraction process. They also include the temperature, speed, and time used in a first centrifugation, and the temperature used in a second centrifugation during the process. The goal of the screening experiment was to identify factors with active linear or quadratic effects and active interaction effects on the amount of extracted pesticides. 

\begin{table}[h!]
    \centering
    \caption{Experimental factors of the extraction experiment with their settings and coded levels.}
    \begin{tabular}{clccc} \toprule
 \multirow{2}{*}{Label}    & \multirow{2}{*}{Factor}  & \multicolumn{3}{c}{Coded Levels} \\ \cmidrule{3-5}
     &  & $-1$ & 0 & 1 \\ \midrule
$X_1$ & Agitation time (min) & 20 & 30 & 40 \\
$X_2$ & Shaking time 1 (min) & 2 & 5 & 8 \\
$X_3$ & Centrifuge temperature 1 ($^\circ$C) & 16 & 20 & 24 \\
$X_4$ & Centrifuge speed (g) & 3743 & 6654 & 10397 \\
$X_5$ & Centrifuge time (min) & 3 & 5 & 7 \\
$X_6$ & Shaking time 2 (min) & 2 & 5 & 8 \\
$X_7$ & Centrifuge temperature 2 ($^\circ$C) & 16 & 20 & 24 \\ \bottomrule
    \end{tabular}
    \label{tab:factors}
\end{table}

% OMARS designs

\cite{Maestronietal2018} performed the experiment using a completely randomized factorial design that is now known as an Orthogonal Minimally Aliased Response Surface (OMARS) design \citep{NunezAresGoos2020,Goos2025}. An OMARS design tests each factor at three levels, permitting the study of the individual quadratic effects of the factors. In addition, the linear effects are not aliased with each other nor with the quadratic effects and two-factor interactions in the design. Therefore, if we assume that third- and higher-order effects are negligible, the least squares estimates of the linear effects are unbiased.  

\cite{NunezAresGoos2020} introduced an enumeration algorithm to obtain a complete catalog of OMARS designs with up to five factors and 44 runs, and a partial catalog of designs with six and seven factors and up to 70 runs. The algorithm has been extended to produce OMARS designs with additional two-level factors \citep{NunezAresetal2023a}. However, it is computationally infeasible to generate OMARS designs with more than seven three-level factors using the algorithm of \cite{NunezAresGoos2020}.

At the time of the extraction experiment, the enumeration algorithm of \cite{NunezAresGoos2020} was not available, because the OMARS designs, as we now know them, had not yet been discovered. \cite{Maestronietal2018} constructed their OMARS design by concatenating two copies of the Definitive Screening Design \citep[DSD;][]{Jones2011} with 17 runs and seven factors in \cite{,Xiao2012}. Nowadays, a DSD is considered a special case of an OMARS design. It is also now known that a DSD can be constructed by folding over a conference design \citep{Schoenetal2022} and adding a center run in which all factors are set to their middle level. The minimum run size to study an even number of factors is twice the number of factors plus one, making a DSD one of the smallest OMARS designs for this case. The minimum run size for a DSD with an odd number of factors is twice the number of factors plus three. \cite{Schoenetal2022} provided a complete catalog of conference designs and DSDs with up to 24 factors, which was also not available when the extraction experiment was designed.

\cite{Maestronietal2018} concatenated the copies of the 17-run DSD following the methodology of \cite{Vazquezetal2019}, who concatenated two two-level orthogonal arrays \citep{hedayat1999} to produce attractive two-level orthogonal designs with large run sizes. In one of the copies of the DSD, the order of the factor columns was changed and specific factor columns were folded over using the CC/VNS algorithm of \cite{Vazquezetal2019}, which we explain later in the paper. Table~\ref{tab:OMARS_design} shows the 7-factor 34-run OMARS design constructed by \cite{Maestronietal2018}. The design has less aliasing among second-order effects than the 17-run 7-factor DSD. It is also more attractive than other 7-factor designs that were available in the literature at the time the experiment was designed. These include face-centered central composite designs, Box-Behnken designs, fractional Box-Behnken designs \citep{edwards2011fractional}, orthogonal arrays \citep{Cheng2001,Xu2004}, and orthogonal array composite designs \citep{xu2014combining,Zhouxu2017}. These benchmark designs have more runs or provide linear effects that are aliased with two-factor interactions. These reasons motivated \cite{Maestronietal2018} to use the design in Table~\ref{tab:OMARS_design}.

\begin{table}[]
    \centering
        \caption{OMARS design with seven factors and 34 runs used in the extraction experiment.}
    \begin{tabular}{lrrrrrrrclrrrrrrr} \toprule
Run    & $X_1$ & $X_2$ & $X_3$ & $X_4$ & $X_5$ & $X_6$ & $X_7$ & & Run & $X_1$ & $X_2$ & $X_3$ & $X_4$ & $X_5$ & $X_6$ & $X_7$ \\ \cmidrule{1-8} \cmidrule{10-17}
1	&	0	&	1	&	1	&	1	&	1	&	1	&	1	& &	18	&	$-1$	&	1	&	$-1$	&	1	&	0	&	1	&	1	\\
2	&	0	&	1	&	1	&	$-1$	&	$-1$	&	1	&	1	& &	19	&	1	&	$-1$	&	$-1$	&	1	&	0	&	$-1$	&	1	\\
3	&	0	&	$-1$	&	$-1$	&	1	&	1	&	$-1$	&	$-1$	& &	20	&	1	&	$-1$	&	1	&	$-1$	&	0	&	$-1$	&	$-1$	\\
4	&	0	&	$-1$	&	$-1$	&	$-1$	&	$-1$	&	$-1$	&	$-1$	& &	21	&	$-1$	&	$-1$	&	1	&	1	&	$-1$	&	0	&	$-1$	\\
5	&	$-1$	&	0	&	$-1$	&	$-1$	&	$-1$	&	1	&	1	& &	22	&	1	&	1	&	$-1$	&	$-1$	&	$-1$	&	0	&	$-1$	\\
6	&	$-1$	&	0	&	1	&	$-1$	&	1	&	1	&	$-1$	& &	23	&	1	&	1	&	$-1$	&	$-1$	&	1	&	0	&	1	\\
7	&	1	&	0	&	$-1$	&	1	&	$-1$	&	$-1$	&	1	& &	24	&	$-1$	&	$-1$	&	1	&	1	&	1	&	0	&	1	\\
8	&	1	&	0	&	1	&	1	&	1	&	$-1$	&	$-1$	& &	25	&	$-1$	&	$-1$	&	$-1$	&	1	&	1	&	1	&	0	\\
9	&	$-1$	&	1	&	0	&	1	&	$-1$	&	$-1$	&	1	& &	26	&	1	&	$-1$	&	1	&	1	&	$-1$	&	1	&	0	\\
10	&	1	&	1	&	0	&	1	&	1	&	1	&	$-1$	& &	27	&	1	&	1	&	1	&	$-1$	&	$-1$	&	$-1$	&	0	\\
11	&	1	&	$-1$	&	0	&	$-1$	&	1	&	1	&	$-1$	& &	28	&	$-1$	&	1	&	$-1$	&	$-1$	&	1	&	$-1$	&	0	\\
12	&	$-1$	&	$-1$	&	0	&	$-1$	&	$-1$	&	$-1$	&	1	& &	29	&	$-1$	&	$-1$	&	1	&	$-1$	&	1	&	$-1$	&	1	\\
13	&	$-1$	&	1	&	$-1$	&	0	&	1	&	$-1$	&	$-1$	& &	30	&	1	&	$-1$	&	$-1$	&	$-1$	&	1	&	1	&	1	\\
14	&	$-1$	&	$-1$	&	$-1$	&	0	&	$-1$	&	1	&	$-1$	& &	31	&	1	&	1	&	$-1$	&	1	&	$-1$	&	1	&	$-1$	\\
15	&	1	&	$-1$	&	1	&	0	&	$-1$	&	1	&	1	& &	32	&	$-1$	&	1	&	1	&	1	&	$-1$	&	$-1$	&	$-1$	\\
16	&	1	&	1	&	1	&	0	&	1	&	$-1$	&	1	& &	33	&	0	&	0	&	0	&	0	&	0	&	0	&	0	\\
17	&	$-1$	&	1	&	1	&	$-1$	&	0	&	1	&	$-1$	& &	34	&	0	&	0	&	0	&	0	&	0	&	0	&	0	\\ \bottomrule
    \end{tabular}
    \label{tab:OMARS_design}
\end{table}

% Problem. Constructing large OMARS
\reva{Although \cite{Maestronietal2018} were the first to construct an OMARS design by concatenating two DSDs, they did not perform a systematic study of the statistical properties of this type of concatenated design. In this paper, we characterize this type of design in full and provide novel concatenated OMARS designs with more than seven factors. Specifically, we construct OMARS designs by concatenating two equally-sized DSDs with up to 20 factors, which we obtained from the catalog of \cite{Schoenetal2022}. We concatenate two copies of a single DSD and, in some cases, two non-isomorphic DSDs; see Section~\ref{sec:construction}. We refer to the resulting concatenated designs as cOMARS designs. 

Although OMARS designs can be used as response surface designs, we tailor our cOMARS designs for use in screening experiments. We show that cOMARS designs have certain statistical properties that do not depend on how the DSDs are concatenated. They include the efficiencies for estimating models with linear and quadratic effects and the aliasing between two quadratic effects. Properties that depend on the concatenation of the DSDs include the aliasing between a quadratic effect and a two-factor interaction, and between pairs of two-factor interactions. We characterize the aliasing using analytical formulas or theoretical results that are new to the literature. Moreover, we minimize the aliasing among the second-order effects by adapting the CC/VNS algorithm of \cite{Vazquezetal2019} to concatenate three-level designs.}

The main merit of our work is that our modified CC/VNS algorithm constructs cOMARS designs with numbers of runs and factors that cannot be handled by the enumeration approach of \cite{NunezAresGoos2020} and for which there are no benchmarks in the literature. However, to demonstrate that the modified CC/VNS algorithm works well, we compare the performance of our cOMARS designs with benchmark DSDs and OMARS designs whenever these are available. \reva{Using simulations, we also compare the 7-factor cOMARS design in Table~\ref{tab:OMARS_design} with non-orthogonal designs that do not impose restrictions in terms of the aliasing between the first- and second-order effects. We demonstrate that our cOMARS design competes with these designs in terms of the power, type-I error rate, false discovery rate, and estimation efficiency for problems with several first-order effects and a limited number of second-order effects.}

The remainder of the paper is organized as follows. In Section~\ref{sec:construction}, we introduce the basic construction of a cOMARS design and, in Section~\ref{sec:properties}, we discuss its statistical properties. In Section~\ref{sec:algorithmic}, we present the CC/VNS algorithm of \cite{Vazquezetal2019} and its adaptation to optimize cOMARS designs. In Section~\ref{sec:results}, we present a collection of cOMARS designs with seven to 20 factors and, in Section~\ref{sec:Examples}, we compare them with DSDs and OMARS designs in the literature. In Section~\ref{sec:non_orthogonal}, we compare the 7-factor cOMARS design in Table~\ref{tab:OMARS_design} with non-orthogonal designs using simulations. In Section~\ref{sec:discussion}, we end the article with a discussion. The supplementary material of the paper contains the derivations of the properties of the cOMARS designs\reva{, additional simulation results}, a Matlab implementation of our adapted CC/VNS algorithm, the collection of cOMARS designs with seven to 20 factors, and \reva{R and Python} code to reproduce our simulation study.

\section{Design Construction} \label{sec:construction}

We first introduce the parent designs for our cOMARS designs. Next, we present the basic construction of a cOMARS design. 

\subsection{Parent designs}

We use $m$ to denote the number of factors and define a \textit{parent} DSD as a DSD without the center run. We construct a parent DSD by folding over a conference design with $m$ factors and $n$ runs \citep{Schoenetal2022}. Formally, a conference design is an $n \times m$ matrix with orthogonal columns, one zero in each column, at most one zero in each row, and $-1$ and $+1$ entries elsewhere, where $n$ is a multiple of two larger than or equal to $m$. If $n = m$, a conference design is also known as a conference matrix \citep{Colbourn2006}. For a given number of factors, there are conference designs with different values of $n$ \citep{Schoenetal2022}. To limit the run size of our cOMARS designs, we restrict our attention to parent DSDs constructed using the smallest conference designs available. Specifically, if $m$ is odd, we obtain a parent DSD with $2n$ runs by folding over an $m$-factor conference design with $n = m+1$ runs. If $m$ is even, we obtain a parent DSD with $2n$ runs by folding over an $m$-factor conference design with $n=m$ runs. 

We obtain conference designs with $m \leq 20$ from the catalog of \citet{Schoenetal2022}. For $m$ equal to 19 and 20, the catalog has two non-isomorphic conference designs. Two conference designs are non-isomorphic if one cannot be obtained from the other by permuting rows or columns, or folding over one or more columns. Non-isomorphic  conference designs produce non-isomorphic DSDs. So, for $m$ equal to 19 and 20, there are two different DSDs to consider as parent designs. For the other numbers of factors, the catalog includes only one conference design, as a result of which there is only one DSD to consider as a parent design.

We denote the two parent DSDs with $m$ factors by $\mathbf{D}_1$ and $\mathbf{D}_2$. For $m<19$, $\mathbf{D}_1=\mathbf{D}_2$ in our construction because only one DSD is available for use as a parent design. For $m$ equal to 19 and 20, we can study the situation where $\mathbf{D}_1\ne\mathbf{D}_2$, due to the existence of two non-isomorphic DSDs. In the remainder of this paper, we refer to $\mathbf{D}_1$ and $\mathbf{D}_2$ as the upper and lower parent DSD, respectively.

\subsection{Basic construction}

The basic construction of a cOMARS design $\mathbf{C}$ is
\begin{equation*}\label{eq:comars}
    \mathbf{C} = \begin{pmatrix}
         \mathbf{D}_1  \\ \mathbf{D}_2 \\ \mathbf{0}_{n_0 \times m} 
    \end{pmatrix}, 
\end{equation*}
\noindent where $n_0$ is the desired number of center runs and $\mathbf{0}_{n_0 \times m}$ is an $n_0 \times m$ matrix of zeros. For an even number of factors, the run size of the cOMARS design is $4 m + n_0$. For an odd number of factors, the run size of the design is $4 (m + 1) + n_0$. 

\cite{NunezAresGoos2020} classify OMARS designs in terms of the number of zeros in the columns of a model matrix corresponding to a model with linear effects and two-factor interaction effects, not counting the center runs. Design $\mathbf{C}$ belongs to the class of OMARS designs with four zeros in each linear effect column of the model matrix and eight zeros in each two-factor interaction column, without the $n_0$ center runs. \reva{This design also belongs to the class of uniform-precision OMARS designs \citep{Goos2025}, which have an equal number of zeros in each linear effect column and allow all linear effects to be estimated with the same precision. This is in contrast with non-uniform-precision OMARS designs that have unequal numbers of zeros in their linear effect columns.}

\section{Statistical Properties} \label{sec:properties}

In this section, we present the statistical properties of a cOMARS design. Specifically, we show that the degree of aliasing between quadratic effects in a cOMARS design can be computed using analytical formulas that do not depend on the parent DSDs. The aliasing between other pairs of second-order effects does depend on the parent DSDs; for its characterization, we derive analytical formulas and a theoretical result. Supplementary Section~A contains the derivation of the theoretical result and all formulas reported in this section. It also shows additional results on the estimation efficiencies for models with linear and quadratic effects\reva{, quantified using the D criterion \citep{Atkinson2007} and standard errors of least squares estimates.}

In the remainder of this section, we consider a parent DSD obtained from a conference design with $m > 4$ and $n = m + [m \Mod 2]$. So, we assume the conference design has five or more factors and its run size is a multiple of two.

\subsection{Correlations between selected second-order effect columns} \label{sec:correlations_specific_SOcols}

We consider a model matrix with columns corresponding to linear effects, quadratic effects, and two-factor interactions of the $m$ factors. We measure the aliasing between two effects by the absolute correlation between the corresponding columns in the model matrix. The higher the absolute correlation value, the higher the degree of aliasing between the effects. An absolute correlation of one implies that the effects are fully aliased, while an absolute correlation of zero implies that the effects are not aliased at all. An absolute correlation between zero and one means that the effects are partially aliased.

\subsubsection{Correlation between two quadratic effect columns} \label{sec:correlations_QEs}

Regardless of the parent DSDs, the correlation between the columns corresponding to the quadratic effects of any two factors $i$ and $j$ in an $m$-factor cOMARS design with $4n + n_0$ runs only depends on $n_0$ and $n$. More specifically, this correlation is
\begin{equation} 
r_{ii,jj} = \frac{n_0 (n - 2) - 4}{(n - 1)(n_0 + 4)}, \nonumber
\end{equation}
\noindent and increases with $m$ because $n = m$ or $n = m+1$ depending on whether $m$ is even or odd. 

\begin{figure}
    \centering
    \begin{subfigure}[b]{0.48\textwidth}
        \includegraphics[width=\textwidth]{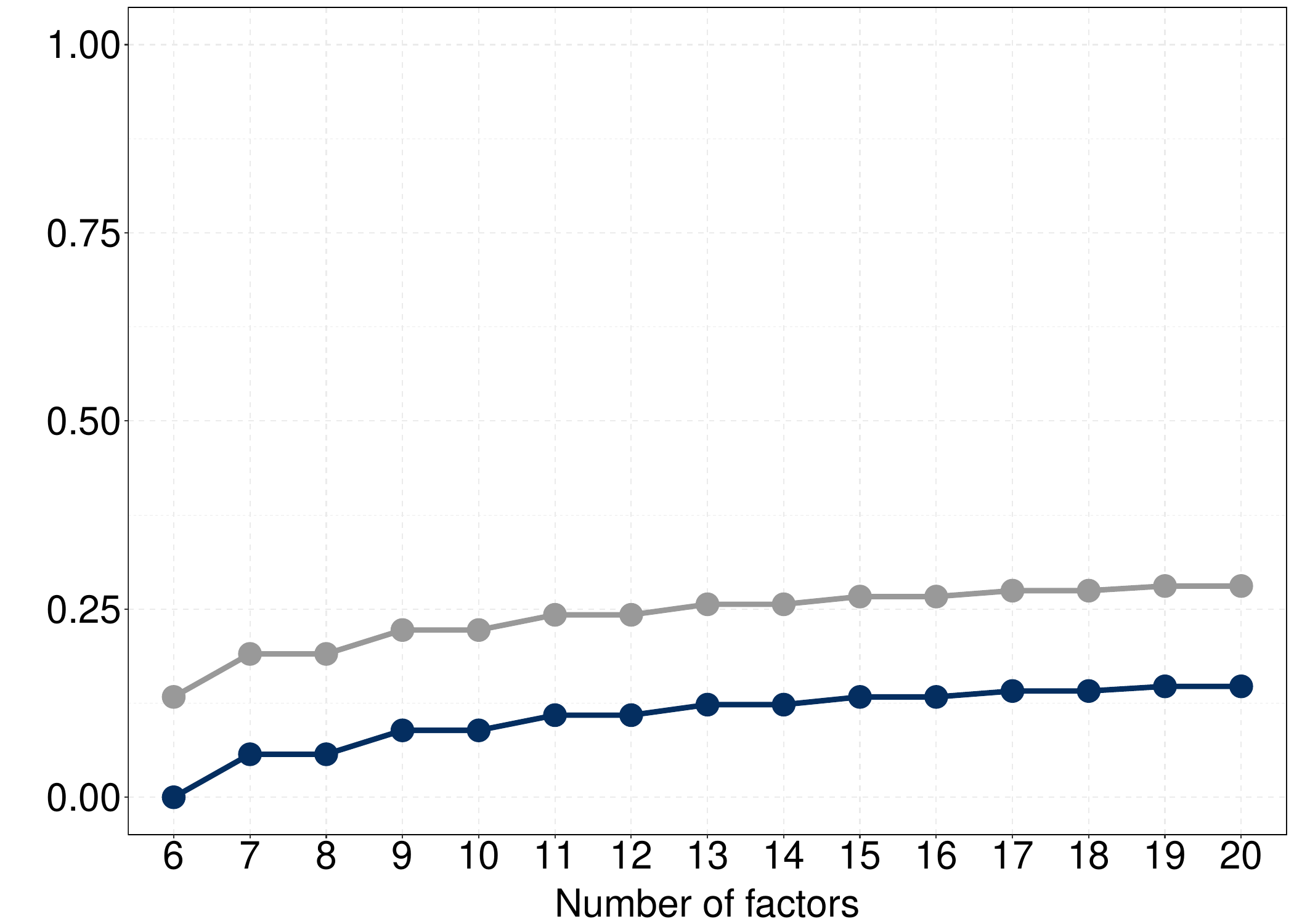}
        \caption{$n_0 = 1$}
        \label{fig:pqen1}
    \end{subfigure}
    ~ %add desired spacing between images, e. g. ~, \quad, \qquad, \hfill etc. 
    \begin{subfigure}[b]{0.48\textwidth}
        \includegraphics[width=\textwidth]{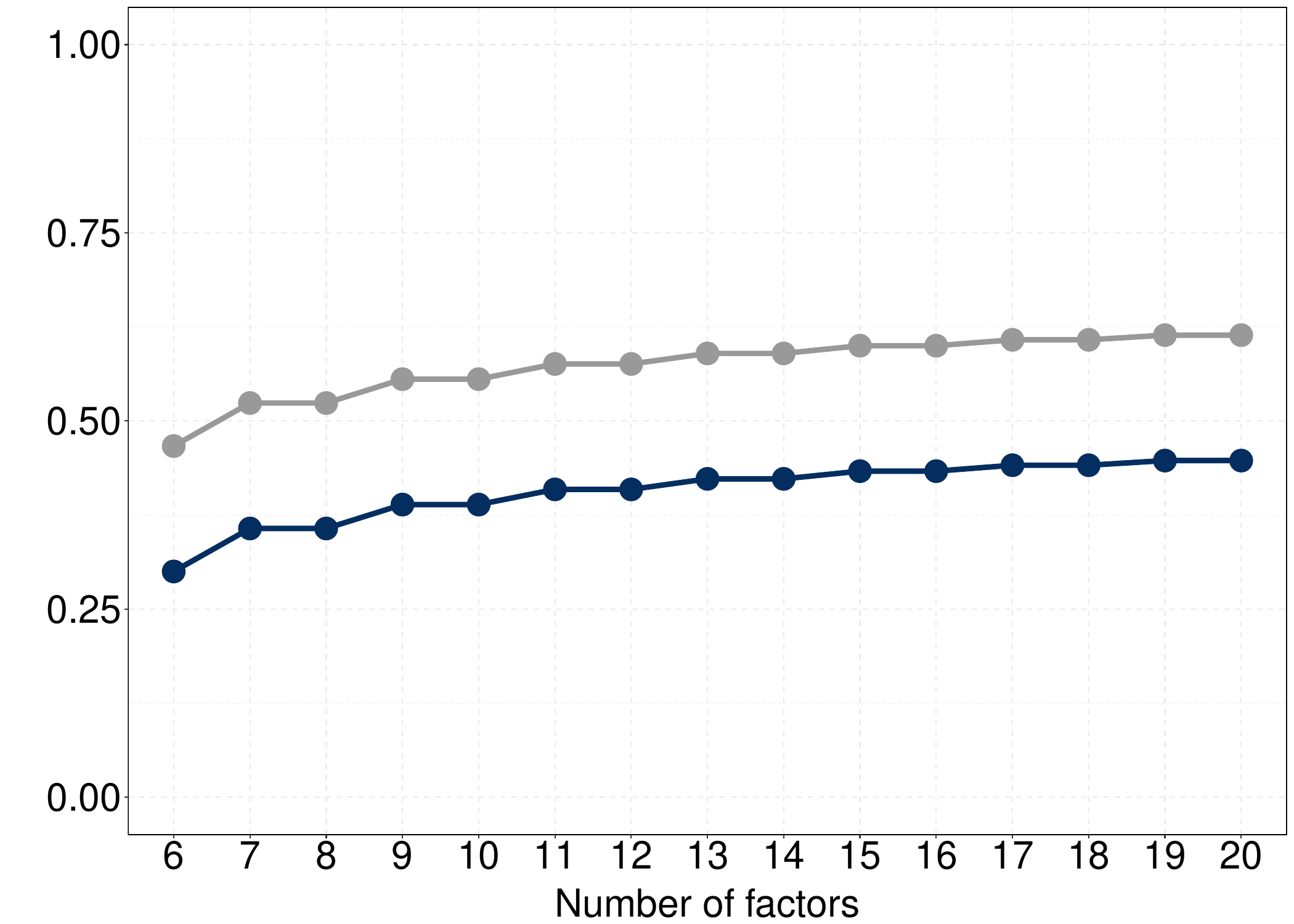}
        \caption{$n_0 = 4$}
        \label{fig:pqen2}
    \end{subfigure}
    \caption{Correlations between two quadratic effect columns of the model matrix for DSDs (gray) and cOMARS designs (blue). The online version of this figure is in color.}\label{fig:peq}
\end{figure}

Figure~\ref{fig:peq} compares the correlation between two quadratic effect columns for DSDs and cOMARS designs with six to 20 factors and one or four center runs. The analytical formula for this correlation in DSDs is $[n_0 (n - 2) - 2]/(n-1)(n_0 + 2)$ \citep{Georgiou2014}. Figure~\ref{fig:peq} shows that the cOMARS designs have a smaller correlation between the quadratic effect columns than the DSDs. For both designs, the correlation increases with the number of \reva{factors.} 

\reva{The correlation between quadratic effect columns also increases with the number of center runs. Actually, for cOMARS designs, the correlation converges to $n_0 / (n_0 + 4)$ when $m$ tends to infinity. So, if $n_0$ is equal to one or four, for example, the limit of this correlation is 0.2 or 0.5, respectively. This means that adding center runs to a cOMARS design increases the aliasing between quadratic effects. However, in supplementary Section A.2, we show that, in spite of the increased aliasing, adding center runs increases the efficiency of estimating these effects in a model with the intercept, all linear effects, and all quadratic effects of the $m$ factors. This is because, for a fixed value of $m$, increasing the number of center runs reduces the standard errors of the least squares estimates of the quadratic effects' coefficients.}

\subsubsection{Correlation between a quadratic effect and a two-factor interaction column}

For DSDs, the correlation between columns that correspond to a quadratic effect and a two-factor interaction when they share a factor is zero \citep{Jones2011,Vazquezetal2019}. The same is true for a cOMARS design because they are constructed by concatenating two DSDs. However, unlike correlations in DSDs, the correlation between a quadratic effect column and a two-factor interaction column when they do not share a factor may be zero in cOMARS designs. We denote the absolute value of this correlation by $r_{ii,jk}$. Depending on how the parent DSDs are concatenated, it can take two possible values in an $m$-factor cOMARS design with $4n + n_0$ runs: 
\begin{equation} 
%r_{ii,jk} = 
\sqrt{\frac{4n + n_0}{(n_0 + 4)(n - 1)(n -2)}} \mbox{ or } 0. \nonumber
\end{equation}
\noindent The non-zero value of $r_{ii,jk}$ decreases with $n$ or $m$. So, the most severe aliasing between a quadratic effect and an interaction that do not share a factor decreases with the number of factors. For DSDs, the correlation $r_{ii,jk}$ is always different from zero and equal to $[(2n + n_0)/(n_0 + 2)(n - 1)(n - 2)]^{1/2}$ \citep{Georgiou2014}.

\begin{figure}
    \centering
    \begin{subfigure}[b]{0.48\textwidth}
        \includegraphics[width=\textwidth]{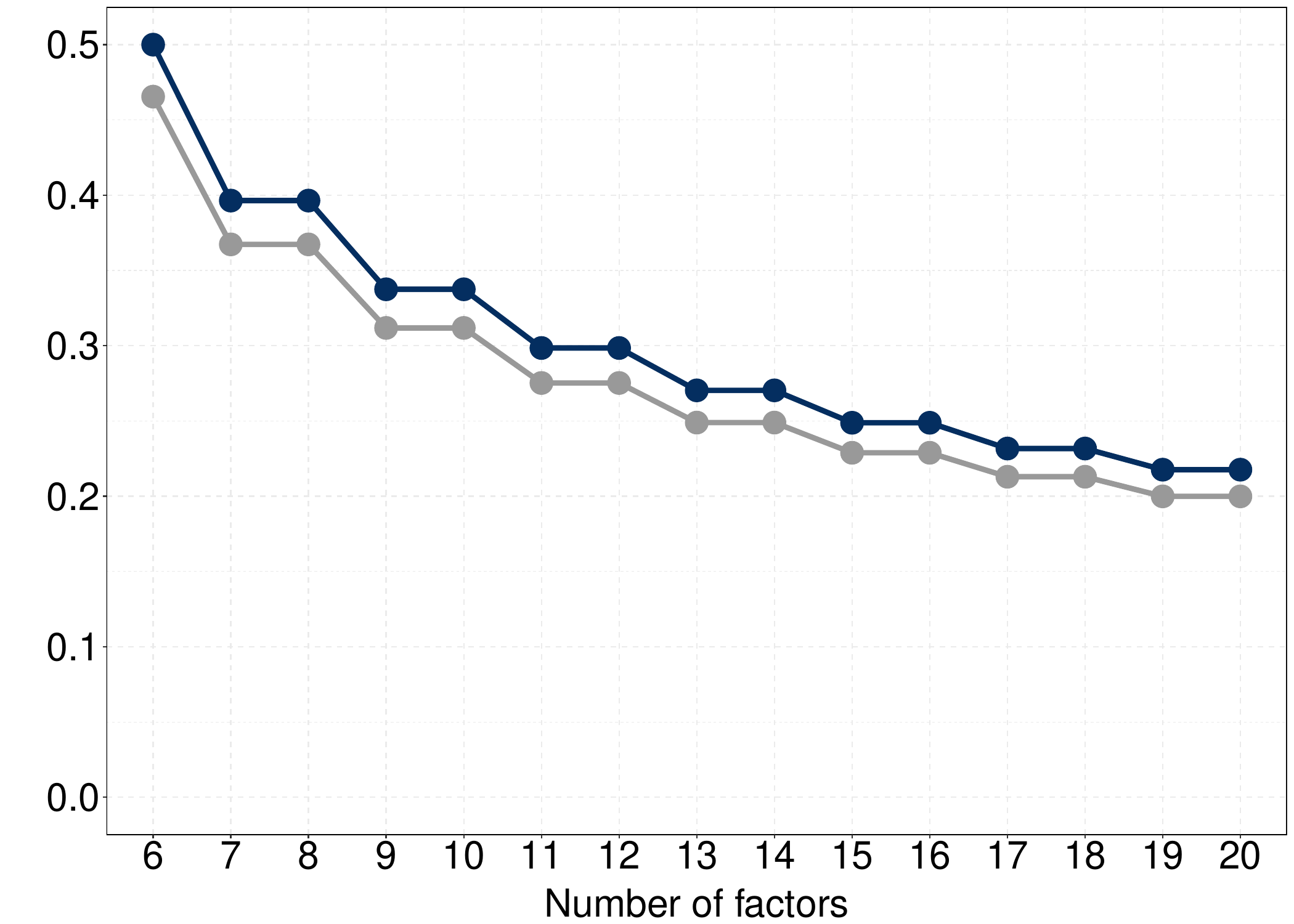}
        \caption{$n_0 = 1$}
        \label{fig:twofin1}
    \end{subfigure}
    ~ %add desired spacing between images, e. g. ~, \quad, \qquad, \hfill etc. 
    \begin{subfigure}[b]{0.48\textwidth}
        \includegraphics[width=\textwidth]{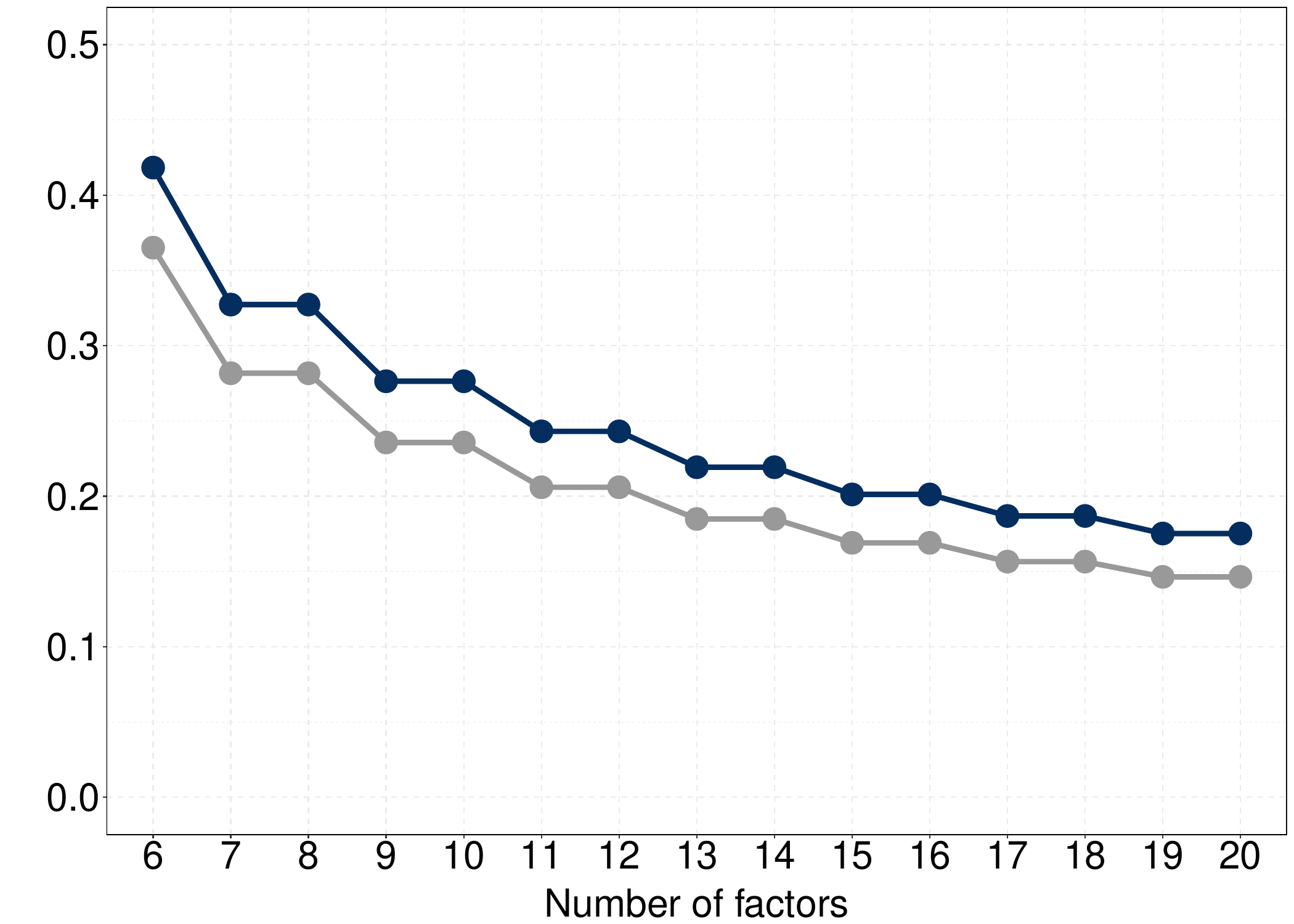}
        \caption{$n_0 = 4$}
        \label{fig:twofin2}
    \end{subfigure}
    \caption{Maximum absolute correlations between a quadratic effect column and a two-factor interaction column for DSDs (gray) and cOMARS designs (blue). The online version of this figure is in color.}\label{fig:twofi}
\end{figure}

Figure \ref{fig:twofi} compares the maximum absolute correlation between the quadratic effect and two-factor interaction columns for DSDs and cOMARS designs. It shows that DSDs provide a lower maximum absolute correlation between these types of columns than cOMARS designs. However, for both designs, the maximum absolute correlation tends to zero as the number of factors increases.  

\subsubsection{Correlation between two-factor interaction columns that share a factor}

In a cOMARS design, the absolute correlation between pairs of two-factor interaction columns that share a factor, which we denote as $r_{ij,ik}$, has two possible values: 
\begin{equation*} 
%r_{ij,ik} = 
\frac{1}{n - 2} \mbox{ or } 0. 
\end{equation*}
\noindent The value of $r_{ij,ik}$ depends on how the parent DSDs are concatenated, but its maximum absolute value decreases with $n$ or $m$. For a DSD, the value of $r_{ij,ik}$ is equal to $1/(n-2)$ \citep{Vazquezetal2020}. So, the maximum absolute correlation between these pairs of columns is similar for DSDs and cOMARS designs. However, the latter can have correlations equal to zero.  

In supplementary Section~A.5, we show that $r_{ij,ik}$ is proportional to $r_{ii,jk}$. Therefore, a cOMARS design with a small correlation between pairs of interaction columns that share a factor also has a small correlation between the quadratic effect columns and the interaction columns. We use this fact when optimizing cOMARS designs using the CC/VNS algorithm in Section~\ref{sec:algorithmic}.

\subsection{Correlations between two-factor interaction columns that do not share a factor} \label{sec:corr_four_factor}

We denote the absolute correlation between pairs of two-factor interaction columns that do not share a factor by $r_{ij,lk}$. For DSDs, \cite{Schoenetal2019} show that $r_{ij,lk}$ is proportional to the $J_4$-characteristic, defined as the absolute value of the sum of the elements of a four-factor interaction column. The smaller the $J_4$-characteristic, the smaller the absolute correlation between three pairs of two-factor interactions involving four different factors. If we denote the four factors involved in the two interactions by $i$, $j$, $k$, and $l$, the pairs of two-factor interactions can be represented as $(ij, lk)$, $(il, jk)$, and $(ik, jl)$. 

For a cOMARS design, $r_{ij,lk}$ is also obtained from the $J_4$-characteristic. In fact, it is calculated by dividing the $J_4$ characteristic by the number of non-zero entries in the corresponding four-factor interaction column of the design. In supplementary Section~A.6, we prove the following result concerning the possible values of the $J_4$-characteristic for a cOMARS design. 

\begin{theorem} \label{theo:J4}
For a cOMARS design constructed by concatenating two $m$-factor parent DSDs with $2n$ runs:
\begin{itemize}
    \item If $n \equiv 0 \Mod 4$, the possible values of the $J_4$-characteristic are $4n - 8 \lambda$ for $\lambda = 2, 3, \ldots, n/2$.
    \item If $n \equiv 2 \Mod 4$, the possible values of the $J_4$-characteristic are $16 \lambda$ and $4n - 16(\lambda + 1)$ for $\lambda = 0, 1, \ldots, (n-6)/4$.
\end{itemize}
\end{theorem}
\noindent This result holds for parent DSDs constructed from a single conference design or from two non-isomorphic conference designs.  

%Theorem~\ref{theo:J4} 
The possible values of the absolute correlation between pairs of interaction columns involving four factors in a cOMARS design follow from our result. For parent DSDs obtained from conference designs with a run size that is a multiple of four, $r_{ij,lk}$ has $n/2 - 1$ possible values: 
\begin{equation} \label{eq:2FI2FI_even}
\frac{n - 2 \lambda}{n - 2},
\end{equation}
\noindent for $\lambda = 2, 3, \ldots, n/2$; see supplementary Section~A.7. For parent DSDs generated from conference designs with a run size that is two more than a multiple of four, $r_{ij,lk}$ has $(n-6)/4+1$ possible values:
\begin{equation} \label{eq:2FI2FI_odd}
\frac{4 \lambda}{n - 2} \mbox{ or } \frac{n - 4(\lambda + 1)}{n - 2},
\end{equation}
\noindent for $\lambda = 0, 1, \ldots, (n-6)/4$. For either case, the actual values of the absolute correlations depend on the parent DSDs and on the way they are concatenated. 

%%%%%%%%%%%%%%%%%%%%%%%%%%

\section{Algorithmic Improvement of Concatenated Designs using the CC/VNS algorithm}
\label{sec:algorithmic}

In a cOMARS design, the correlations between columns corresponding to a quadratic effect and an interaction and pairs of interactions depend on the parent DSDs and on the way in which they are concatenated. To minimize these correlations, we adapt the CC/VNS algorithm of \cite{Vazquezetal2019} to concatenate DSDs. 

Originally, the CC/VNS algorithm constructs attractive designs by concatenating two two-level orthogonal arrays of strength three \citep{hedayat1999}, which possess similar aliasing properties as cOMARS designs. That is, the linear effects are not aliased with each other nor with two-factor interactions in these orthogonal arrays. To optimize the concatenated design, the algorithm performs column permutations and folds over subsets of columns in one of the parent two-level orthogonal arrays. In this way, it avoids a complete enumeration of all possible column permutations and foldovers of all subsets of columns in the parent design, which is computationally infeasible when the number of factors is large. Using the algorithm, \cite{Vazquezetal2019,Vazquezetal2022} constructed two-level orthogonal designs with up to 33 factors and 128 runs that have limited aliasing between the two-factor interactions. The good performance of the CC/VNS algorithm and the similarity of its input designs to our parent DSDs make it an attractive tool for optimizing cOMARS designs. 

\subsection{Objective function} \label{sec:objective_functions}

In the original CC/VNS algorithm, there are two objective functions that summarize the aliasing between interactions. \reva{The first one concerns the $G_2$-aberration criterion for two-level orthogonal arrays \citep{tang1999minimum}, which is based on the generalized word counts. Of all generalized word counts embedded in this criterion, the generalized word count of length 4, symbolized by $B_4$, is the most relevant for strength-3 orthogonal arrays. This is because $B_4$ is proportional to the sum of squared correlations between pairs of two-factor interaction columns involving four different factors. Pairs that share a factor are not addressed because their correlation is zero in these arrays. So, the lower the value of $B_4$, the smaller the overall aliasing between two-factor interactions. The other word counts of the $G_2$-aberration criterion are irrelevant for strength-3 arrays because they are zero or involve three- and higher-order interactions, which are assumed to be negligible. 

The second objective function concerns the $G$-aberration criterion, which, for two-level strength-3 orthogonal arrays, is based on the $F_4$ vector \citep{deng1999generalized}. This vector contains} the frequencies of each possible value of the correlation between pairs of two-factor interaction columns involving four different factors. Specifically, the objective function is a linear combination of the entries in the $F_4$ vector, in which the entries associated with higher absolute correlations have a higher weight than those associated with lower absolute correlations. Minimizing this objective function sequentially minimizes the most severe aliasing between two-factor interactions.

\reva{Our adaptation of the CC/VNS algorithm has two objective functions inspired by the $B_4$ value and the $F_4$ vector used in the original algorithm. However, our objective functions} account for correlations between pairs of two-factor interaction columns that share a factor as well as pairs that do not share a factor. \reva{This is because, in our cOMARS designs, correlations between pairs of interaction columns that share a factor may be different from zero. The first objective function,} which we call SSQ, is the sum of squared correlations between these pairs of columns. The other objective function, which we call $F$, is a linear combination of the frequencies of the possible absolute values of these correlations, where we use weights similar to those used by \cite{Vazquezetal2019}. Importantly, the development of the objective function $F$ is possible due to Result~\ref{theo:J4} and Equations~\eqref{eq:2FI2FI_even} and \eqref{eq:2FI2FI_odd}. 

Minimizing the SSQ or $F$ function also minimizes the correlations between the quadratic effect and the two-factor interaction columns. This is because these correlations are proportional to those between pairs of interaction columns that share a factor; see Section~\ref{sec:correlations_specific_SOcols}.

\subsection{Overview of the operations of the algorithm}

In our adaptation, the operations of the CC/VNS algorithm on the parent designs are the same as those in \cite{Vazquezetal2019}. That is, the upper parent DSD is fixed and the operations are performed on a starting version of the lower parent DSD, obtained after randomly permuting its columns and folding over a randomly selected subset of its columns. The goal of the algorithm is to minimize the objective function (either SSQ or $F$) set by the user.

The modifications of the starting lower parent DSD are performed by two interconnected algorithms called the column-change (CC) algorithm and the variable neighborhood search (VNS) algorithm. The CC algorithm is a local search algorithm \citep{Michaelwicz2004} that makes small structured changes to the lower parent DSD. Specifically, it folds over a column and exchanges two columns in this design. The VNS algorithm is based on a metaheuristic algorithm of the same name \citep{Hansen2001} and investigates increasingly diverse new versions of the lower parent DSD. The new versions are produced by folding over one or two columns simultaneously or changing the positions of two or three columns in this design. The new versions of the lower parent DSD produced by the VNS algorithm are used as input to the CC algorithm to obtain a better cOMARS design. 

The CC/VNS algorithm terminates its operations when no better cOMARS design has been obtained from the CC and VNS algorithms. Its output is the best cOMARS design found for the two parent DSDs. To increase the likelihood of finding the cOMARS design with the optimal value of the user-selected objective function, we execute the CC/VNS algorithm several times, each time starting from a randomly chosen version of the lower parent DSD. The repetitions of the algorithm can thus be run in parallel on the various cores of a CPU. 

We refer the reader to \cite{Vazquezetal2019,Vazquezetal2022} for a comprehensive explanation of all the details behind the CC/VNS algorithm. Supplementary Section~B provides a comprehensive evaluation of the computational performance of our adapted CC/VNS algorithm to construct cOMARS designs. A Matlab implementation of the adapted algorithm is in the supplementary material.

%---------------------------------------------------------------
\section{cOMARS Designs with 7 to 20 Factors} \label{sec:results}

In this section, we construct a catalog of cOMARS designs with 7--20 factors using the CC/VNS algorithm. First, we describe the setup of the algorithm and the parent DSDs we used. Next, we present the properties of the cOMARS designs in our newly constructed catalog of cOMARS designs. The supplementary material of this article includes our collection of cOMARS designs.

\subsection{Construction setup}

For seven to 18 factors, the parent DSDs are two copies of the DSD obtained by folding over the single $m$-factor conference design in \cite{Schoenetal2022} for $m = 7, \ldots, 18$. If the value of $m$ is even or odd, the run size of the conference design is $m$ or $m+1$, respectively. For 19 and 20 factors, there are two non-isomorphic conference designs which, after folding over, result in two non-isomorphic parent DSDs \citep{Schoenetal2022}. For each of these two numbers of factors, we construct three cOMARS designs by concatenating three pairs of parent DSDs. In two of these pairs, $\mathbf{D}_1$ and $\mathbf{D}_2$ are copies of an individual parent DSD. In the other pair, $\mathbf{D}_1$ and $\mathbf{D}_2$ are the two distinct non-isomorphic DSDs. We report the best of the three resulting cOMARS designs in terms of our objective functions. 

To construct cOMARS designs with seven to 18 factors, we execute the CC/VNS algorithm 100 times to optimize the SSQ and $F$ functions. For the 19- and 20-factor cOMARS designs, we use 10 repetitions only because the algorithm becomes computationally expensive for such large numbers of factors. In any case, in Section~\ref{sec:Examples}, we show that 10 repetitions of the CC/VNS algorithm suffices to obtain attractive cOMARS designs. By convention, our cOMARS designs include a center run that does not affect the aliasing between interactions.

\subsection{A collection of designs}

Table~\ref{tab:collection_designs} shows the correlations between pairs of two-factor interaction columns for the cOMARS designs in our catalog. In Table~\ref{tab:collection_designs}, designs of type `s' and `f' minimize the SSQ and $F$ function, respectively. The table shows the design dimensions in the first two columns and, in bold font, the possible non-zero values for the absolute correlations between pairs of interaction columns. The table also shows the number of pairs of interaction columns with each correlation value. The smallest correlation reported is for pairs of interaction columns that share a factor ($r_{ij,ik}$). The other non-zero values reported are for pairs of interaction columns that do not share a factor ($r_{ij,lk}$). We determined the possible values of these correlations using Result~\ref{theo:J4} and Equations~\eqref{eq:2FI2FI_even} and \eqref{eq:2FI2FI_odd}. The last column of Table~\ref{tab:collection_designs} shows the sum of squared correlations between all pairs of interaction columns. 

Table~\ref{tab:collection_designs} shows that the cOMARS designs of type `f' have a smaller maximum absolute correlation than the corresponding designs of type `s', except for 8, 11, 12, 17, and 18 factors. For these numbers of factors, the cOMARS designs of both types have the same maximum absolute correlation. Regarding the sum of squared correlations between pairs of interaction columns, Table~\ref{tab:collection_designs} shows that the cOMARS designs of type `s' have a smaller sum than the designs of type `f', except for eight and 12 factors in which they match.

Interestingly, for 19 and 20 factors, the cOMARS designs of type `s' are constructed by concatenating two non-isomorphic parent DSDs. Therefore, concatenating non-isomorphic parent DSDs can yield better cOMARS designs than concatenating two copies of a single parent DSD when minimizing the sum of squared correlations between pairs of interaction columns. 

\begingroup

\setlength{\tabcolsep}{4pt} % Default value: 6pt
\renewcommand{\arraystretch}{0.8} % Default value: 1

\begin{table}[h!] \caption{Properties of the cOMARS designs  with seven to 20 factors constructed using our adaptation of the CC/VNS algorithm. The designs indicated by means of an asterisk are constructed from two non-isomorphic parent DSDs.} \label{tab:collection_designs}
\begin{tabular}{rcccccccccccr} 
  \toprule 
Factors	&	Runs	&	Type	&	\multicolumn{9}{l}{Correlations and frequencies}	&	SSQ	\\ \midrule 
	&		&		&	\bf 0.167	&	\bf 0.333	&	\bf 0.667	&		&		&		&		&		&		&		\\
7	&	33	&	s	&	47	&	36	&	6	&		&		&		&		&		&		&	7.972	\\
	&		&	f	&	45	&	72	&	0	&		&		&		&		&		&		&	9.250	\\
8	&	33	&	s, f	&	72	&	144	&	0	&		&		&		&		&		&		&	18.000	\\
	&		&		&	\bf 0.125	&	\bf 0.250	&	\bf 0.500	&	\bf 0.750	&		&		&		&		&		&		\\
9	&	41	&	s	&	114	&	276	&	12	&	0	&		&		&		&		&		&	22.031	\\
	&		&	f	&	108	&	378	&	0	&	0	&		&		&		&		&		&	25.313	\\
10	&	41	&	s	&	160	&	300	&	0	&	30	&		&		&		&		&		&	38.125	\\
 	&	 	&	f	&	220	&	360	&	60	&	0	&		&		&		&		&		&	40.937	\\
	&		&		&	\bf 0.100	&	\bf 0.200	&	\bf 0.400	&	\bf 0.600	&	\bf 0.800	&		&		&		&		&		\\
11	&	49	&	s	&	243	&	432	&	162	&	0	&	0	&		&		&		&		&	45.630	\\
	&		&	f	&	235	&	534	&	153	&	0	&	0	&		&		&		&		&	48.190	\\
12	&	49	&	s, f	&	324	&	684	&	243	&	0	&	0	&		&		&		&		&	69.480	\\
	&		&		&	\bf 0.083	&	\bf 0.167	&	\bf 0.333	&	\bf 0.500	&	\bf 0.667	&	\bf 0.833	&		&		&		&		\\
13	&	57	&	s	&	438	&	951	&	384	&	54	&	0	&	0	&		&		&		&	85.625	\\
	&		&	f	&	444	&	1089	&	486	&	0	&	0	&	0	&		&		&		&	87.333	\\
14	&	57	&	s	&	588	&	1572	&	432	&	99	&	0	&	0	&		&		&		&	120.500	\\
	&		&	f	&	504	&	1533	&	714	&	0	&	0	&	0	&		&		&		&	125.417	\\
	&		&		&	\bf 0.071	&	\bf 0.143	&	\bf 0.286	&	\bf 0.429	&	\bf 0.571	&	\bf 0.714	&	\bf 0.857	&		&		&		\\
15	&	65	&	s	&	673	&	1368	&	624	&	132	&	57	&	12	&	12	&		&		&	140.087	\\
	&		&	f	&	691	&	1500	&	687	&	324	&	0	&	0	&	0	&		&		&	149.729	\\
16	&	65	&	s	&	808	&	2088	&	648	&	312	&	24	&	48	&	0	&		&		&	189.265	\\
	&		&	f	&	856	&	1962	&	906	&	414	&	12	&	0	&	0	&		&		&	198.327	\\
	&		&		&	\bf 0.063	&	\bf 0.125	&	\bf 0.250	&	\bf 0.375	&	\bf 0.500	&	\bf 0.625	&	\bf 0.75	&	\bf 0.875	&		&		\\
17	&	73	&	s	&	1058	&	2589	&	1506	&	549	&	0	&	0	&	0	&	0	&		&	215.914	\\
 	&	 	&	f	&	1004	&	3183	&	1995	&	327	&	0	&	0	&	0	&	0	&		&	224.328	\\
18	&	73	&	s	&	1236	&	3591	&	1854	&	711	&	0	&	0	&	0	&	0	&		&	276.797	\\
	&		&	f	&	1168	&	4074	&	2604	&	450	&	0	&	0	&	0	&	0	&		&	294.250	\\
	&		&		&	\bf 0.055	&	\bf 0.111	&	\bf 0.222	&	\bf 0.333	&	\bf 0.444	&	\bf 0.556	&	\bf 0.667	&	\bf 0.778	&	\bf 0.889	&		\\
19	&	81	&	s$^\star$	&	1463	&	4023	&	3240	&	720	&	105	&	15	&	0	&	0	&	0	&	319.553	\\
	&		&	f	&	1473	&	4137	&	3468	&	915	&	0	&	0	&	0	&	0	&	0	&	328.546	\\
20	&	81	&	s$^\star$	&	1734	&	5037	&	3978	&	960	&	129	&	15	&	0	&	0	&	0	&	400.759	\\
	&		&	f	&	1756	&	5154	&	4272	&	1182	&	0	&	0	&	0	&	0	&	0	&	411.346	\\
   \bottomrule
\end{tabular}
%$^\star$cOMARS design constructed from non-isomorphic parent DSDs.
\end{table}
\endgroup

%%%%%%%%%%%%%%%%%%%%%%%%%%%%%%%%%%%%%%%%%%%%%%%%%%%%%%%
\section{Comparisons with Benchmark OMARS-type Designs}
\label{sec:Examples}

We compare the performance of our cOMARS designs with the DSDs and OMARS designs in the literature. First, we discuss the 7-factor extraction experiment that motivates this article. Next, we compare our 8-factor cOMARS design with the DSDs in \cite{Schoenetal2022} and the OMARS designs in \cite{Hameedetal2023}. After that, we compare our cOMARS designs with nine to 12 factors with the DSDs in \cite{Schoenetal2022}. 

\subsection{Revisiting the extraction experiment}

The goal of the extraction experiment of \cite{Maestronietal2018} was to study the linear, quadratic, and two-factor interaction effects of the seven factors in Table~\ref{tab:factors}. The experiment was conducted using the 7-factor 34-run cOMARS design in Table~\ref{tab:OMARS_design}, which can be obtained by adding a center run to the 7-factor design of \reva{type `f'} in Table~\ref{tab:collection_designs}. We compare the design in Table~\ref{tab:OMARS_design} with comparable DSDs and OMARS designs in \cite{Schoenetal2022} and \cite{NunezAresGoos2020}, respectively. 

The catalog of \cite{Schoenetal2022} has a complete collection of 48 non-isomorphic DSDs with seven factors and 33 runs. From this collection, \cite{Schoenetal2022} obtained the best DSDs in terms of the maximum absolute correlation and the sum of squared correlations between two-factor interaction columns. Here, we use these 7-factor 33-run DSDs with an extra center run so that their run sizes are 34 and match that of our cOMARS design in Table~\ref{tab:OMARS_design}. We refer to the DSD with the smallest absolute correlation as DSD-1 and to the DSD with the smallest sum of squared correlations as DSD-2. 

For seven factors, \cite{NunezAresGoos2020} provide OMARS designs with up to 70 runs, classified by the number of zeros in the linear and interaction effect columns, not counting those in the center runs. We focus on the 18 OMARS designs with seven factors and 32 runs. Following \cite{NunezAresGoos2020}, we label each of these designs as ``$n^{LE}_{0}$-$n^{IE}_0$-$i$'', where $n^{LE}_{0}$ and $n^{IE}_{0}$ are the number of zeros (excluding the center runs) in the linear and interaction effect columns, respectively, and $i$ is the design's identification number. For example, the design labeled `10-16-45' belongs to the class of OMARS designs with 10 and 16 zeroes in each linear and two-factor interaction column, respectively. This design is the 45th design in the series of OMARS designs with $n^{LE}_{0}=10$ and $n^{IE}_0=16$ in the catalog of \cite{NunezAresGoos2020}.  The 18 OMARS designs do not have center runs. So, we add two center runs to each design so that their run sizes are 34. The OMARS design labeled `2-4-61' is a DSD because it has two and four zeros in each linear and interaction effect column, respectively, excluding the center runs. However, it is not isomorphic to DSD-1 nor to DSD-2.

We compare the designs in terms of their absolute correlations between pairs of second-order effect columns. Specifically, we compute the absolute correlations between columns corresponding to two quadratic effects, a quadratic effect and an interaction, and pairs of interactions that share and do not share a factor. Table~\ref{tab:7OMARS} shows the distributions of these correlations for each design, as well as the sum of squared correlations. 

Table~\ref{tab:7OMARS} shows that the maximum absolute correlation between pairs of second-order effect columns of the cOMARS design is smaller than that of the other designs, except for the OMARS design labeled `4-8-62' which has the same maximum absolute correlation as our design. The table also shows that the median absolute correlation between pairs of second-order effect columns is zero for the cOMARS design. This median is smaller than that of the DSDs and the 13 OMARS designs with an $n^{LE}_{0}$ value equal to two or ranging from 10 to 16. The cOMARS design also has a smaller sum of squared correlations between pairs of second-order effect columns than all benchmark designs, except for the DSDs and the OMARS design labeled `4-8-62'.

\begin{table}[h!]
    \centering
    \caption{Sum of squares, maximum, and quartiles of the absolute correlations between all pairs of second-order effect columns for 7-factor 34-run OMARS designs and DSDs. The minimum and $Q_1$ are zero for all designs.}
    \begin{tabular}{lccccclcccc} \toprule 
Design	&	$Q_2$	&	$Q_3$	&	Max.	&	SSQ	& & 	Design	&	$Q_2$	&	$Q_3$	&	Max.	&	SSQ	\\ \midrule 
10-16-45	&	0.125	&	0.250	&	0.625	&	19.490	& & 	16-24-56	&	0.056	&	0.250	&	0.500	&	25.086	\\
10-16-46	&	0.179	&	0.250	&	0.500	&	20.402	& & 	2-4-61	&	0.071	&	0.201	&	0.857	&	16.566	\\
12-20-47	&	0.167	&	0.333	&	0.604	&	18.674	& & 	20-28-57	&	0.000	&	0.359	&	0.500	&	28.383	\\
12-20-48	&	0.167	&	0.333	&	0.604	&	19.689	& & 	20-28-58	&	0.000	&	0.359	&	0.500	&	25.747	\\
12-20-49	&	0.167	&	0.333	&	0.604	&	20.402	& & 	20-28-59	&	0.000	&	0.359	&	0.500	&	23.860	\\
14-20-50	&	0.198	&	0.333	&	0.833	&	28.958	& & 	20-28-60	&	0.000	&	0.359	&	0.500	&	25.747	\\
14-20-51	&	0.198	&	0.333	&	0.833	&	27.884	& & 	4-8-62	&	0.000	&	0.333	&	0.367	&	15.401	\\
14-20-52	&	0.167	&	0.333	&	0.833	&	24.877	& & 	DSD-1	&	0.071	&	0.201	&	0.571	&	15.586	\\
16-24-53	&	0.056	&	0.250	&	0.500	&	22.690	& & 	DSD-2	&	0.071	&	0.201	&	0.857	&	14.607	\\
16-24-54	&	0.243	&	0.250	&	0.500	&	26.311	& & 	cOMARS	&	0.000	&	0.333	&	0.367	&	16.083	\\
16-24-55	&	0.243	&	0.250	&	0.500	&	27.843	& & 		&		&		&		&		\\	
       \bottomrule
    \end{tabular}
    \label{tab:7OMARS}
\end{table}

The OMARS design labeled `4-8-62' has the same number of zeros in the linear and interaction effect columns of the cOMARS design. Moreover, the benchmark design matches the minimum, maximum, and quartiles of the absolute correlations of our design, but the former has a slightly smaller sum of squared correlations between second-order effect columns than the latter. However, a close inspection of the two designs revealed that the cOMARS design has 45 pairs of second-order effect columns with the maximum absolute correlation of 0.367, corresponding to columns involving a quadratic effect and an interaction effect. In contrast, the OMARS design labeled `4-8-62' has 49 pairs of columns with that correlation. 

We conclude that, compared to the DSDs and OMARS designs now in the literature, our cOMARS design in Table~\ref{tab:OMARS_design} remains an attractive option for the extraction experiment in terms of the aliasing between second-order effects. Our design involves less severe aliasing between these effects than these designs.  

%%%%%%%%%%%
\subsection{OMARS designs and DSDs with eight factors}

\cite{Hameedetal2023} introduce a model selection method to analyze data from OMARS designs. They illustrate their method using 8-factor OMARS designs with 27 and 32 runs. To the best of our knowledge, these are the only OMARS designs with more than seven factors in the literature, excluding DSDs, Box-Behnken designs, and face-centered central composite designs. It is thus instructive to compare them with our 8-factor cOMARS designs. 

The OMARS designs of \cite{Hameedetal2023} differ in run size and in two other aspects. First, they differ in the number of zeros in the linear effect columns. Apart from center runs, there are eight zeros in the 27-run design and four zeros in the 32-run design. Second, the 27-run design can be constructed by folding over a specific 13-run 8-factor design and adding a center run. In contrast, the 32-run design cannot be constructed by folding over a 16-run design and does not have a center run. In our comparisons, we exclude the center run from the 8-factor cOMARS design characterized in Table~\ref{tab:collection_designs}, so that its run size matches that of the 32-run OMARS design of \cite{Hameedetal2023}.

To enrich our comparisons, we include comparable 8-factor DSDs from the catalog of \cite{Schoenetal2022}. The catalog has 77 non-isomorphic DSDs with eight factors and 33 runs, two of which are recommended by these authors. One DSD is best in terms of the maximum absolute correlation between pairs of interaction columns. The other DSD is best in terms of the sum of squared correlation between pairs of interaction columns. We exclude the center run from each of these DSDs, so that their run sizes match those of our cOMARS design and the 32-run OMARS design in \cite{Hameedetal2023}. 

Table~\ref{tab:8OMARS} summarizes the absolute correlations between pairs of second-order effect columns for the DSDs and OMARS designs with eight factors. The maximum absolute correlation of our 32-run cOMARS design is smaller than that of all benchmark designs. Moreover, the median absolute correlation and the sum of squared correlations for our cOMARS design are smaller than those for the 27-run OMARS design. However, our cOMARs design has larger values for these summary statistics than those of the 32-run benchmark designs. 

\begin{table}[h!]
    \centering
    \caption{Sum of squares, maximum, and quartiles of the absolute correlations between pairs of second-order effect columns for 8-factor designs with 27 and 32 runs.}
    \begin{tabular}{llcccc} \toprule 
        &   &  \multicolumn{4}{c}{Absolute correlations}  \\ \cmidrule{3-6}
      Reference	&	Runs	&	$Q_2$	&	$Q_3$	&	Max.	&	SSQ	\\ \midrule
\multirow{1}{*}{\cite{Schoenetal2022}}	&	32	&	0.071	&	0.276	&	0.571	&	28.965	\\
	&	32	&	0.071	&	0.276	&	0.857	&	27.496	\\
\multirow{1}{*}{\cite{Hameedetal2023}}	&	27	&	0.167	&	0.333	&	0.500	&	41.167	\\
	&	32	&	0.000	&	0.167	&	0.667	&	28.516	\\
This paper	&	32	&	0.143	&	0.333	&	0.436	&	32.286	\\
       \bottomrule
    \end{tabular}
    \label{tab:8OMARS}
\end{table}

We conclude that none of the designs in Table~\ref{tab:8OMARS} outperforms the others on all criteria and therefore all designs are competitive. The decision of which design to use depends on whether we prefer to minimize the most severe aliasing or the overall aliasing between second-order effects. If the goal is to minimize the most severe aliasing, we recommend our cOMARS design. Otherwise, we recommend the second DSD of \cite{Schoenetal2022} in Table~\ref{tab:8OMARS}. The 32-run OMARS design of \cite{Hameedetal2023} and the first DSD of \cite{Schoenetal2022} provide a compromise between minimizing the most severe aliasing and minimizing the overall aliasing.

\subsection{DSDs with nine to 12 factors} \label{sec:DSDs}

In addition to the 7- and 8-factor DSDs, \cite{Schoenetal2022} recommended specific 41-run DSDs with nine and ten factors, and specific 49-run DSDs with 11 and 12 factors. We compare our 9-, 10-, 11-, and 12-factor cOMARS designs in Table~\ref{tab:collection_designs} with these DSDs in terms of the aliasing between second-order effects.

Table~\ref{tab:DSDs} summarizes the correlations between pairs of two-factor interaction columns (which share or do not share a factor) for the aforementioned  DSDs. The table shows the maximum absolute correlation and the number of pairs of interaction columns with this correlation, as well as the sum of squared correlations between pairs of interaction columns. For nine factors, our cOMARS design of type `f' has a smaller maximum absolute correlation than the corresponding DSD. For 11 factors, our cOMARS designs also have a smaller maximum absolute correlation than one of the DSDs in Table~\ref{tab:DSDs}. However, the other 11-factor DSD has a smaller maximum absolute correlation than our designs. The same is true for the 10- and 12-factor DSDs. For all numbers of factors in Table~\ref{tab:DSDs}, the recommended DSDs have a smaller sum of squared correlations than our cOMARS designs.  

\begin{table}[h!] 
\begin{center}
\caption{Correlation between two-factor interaction columns in DSDs with nine to 12 factors.} \label{tab:DSDs}
\begin{tabular}{rcccr} 
  \toprule
  Factors	&	Runs	&	Max. Corr.	&	\# Pairs with Max. Corr.	&	SSQ	\\ \midrule 
%7	&	33	&	0.571	&	12	&	7.393	\\
%	&		&	0.857	&	6	&	6.413	\\
%8	&	33	&	0.571	&	30	&	16.041	\\
%	&		&	0.857	&	3	&	14.571	\\
9	&	41	&	0.444	&	36	&	20.923	\\
10	&	41	&	0.444	&	60	&	35.185	\\
11	&	49	&	0.364	&	252	&	43.866	\\
	&		&	0.545	&	3	&	43.469	\\
12	&	49	&	0.364	&	378	&	65.628	\\
   \bottomrule
\end{tabular}
\end{center}
\end{table}

Our cOMARS designs with nine to 12 factors are more attractive than the DSDs in Table~\ref{tab:DSDs} when we include quadratic effects in the analysis. First, our designs have a higher D-efficiency for estimating the model with the intercept, all linear effects, and all quadratic effects. For this model, the D-efficiency of a cOMARS design relative to a DSD is higher than 171\%. So, the cOMARS designs are at least 71\% better than the DSDs; see supplementary Section~A.2. Second, the correlations between two quadratic effect columns of our cOMARS designs are lower than those of the DSDs. Specifically, the correlation between any two quadratic effect columns is 0.281 for the 9- and 10-factor DSDs, and 0.289 for the 11- and 12-factor DSDs. In contrast, the correlations are 0.089 for the 9- and 10-factor cOMARS designs, and 0.109 for the 11- and 12-factor cOMARS designs. 

\section{Comparisons with Non-Orthogonal Designs}\label{sec:non_orthogonal}

In the previous section, we compared the performance of our cOMARS designs with alternative DSDs and OMARS designs in terms of the aliasing between second-order effects. We showed that our cOMARS designs are competitive or better than these benchmark designs. However, in the experimental design literature, there are non-orthogonal designs that do not impose restrictions in terms of the aliasing between first- and second-order effects, as do all DSDs and OMARS designs. Important representatives of non-orthogonal designs are D-optimal designs \citep{goos2011optimal} and Bayesian D-optimal designs \citep{DuMouchel1994}. It is thus instructive to compare our cOMARS designs with these non-orthogonal benchmarks. 

However, it would be unfair to compare non-orthogonal designs with cOMARS designs in terms of the aliasing between their effects. This is because non-orthogonal designs may have linear effects that are aliased with each other and with second-order effects. Therefore, we compare cOMARS designs with non-orthogonal designs using a simulation study inspired by the 7-factor extraction experiment of \cite{Maestronietal2018}. In what follows, we describe the non-orthogonal designs, data analysis methods, simulation protocol, and results.

\subsection{Non-orthogonal designs} \label{sec:nonorthogonal_designs}

A D-optimal design \citep{goos2011optimal} maximizes the efficiency to estimate a \revb{regression} model by maximizing the determinant of the design's information matrix for this model. It is available for any run size that is at least as large as the number of coefficients in the model. For the 7-factor experiment of \cite{Maestronietal2018}, the model under study has 36 coefficients: one intercept, seven linear effects, 21 two-factor interactions, and seven quadratic effects. Therefore, we construct a 36-run 7-factor D-optimal design using the JMP v18 software with 1,000 iterations for its coordinate-exchange algorithm \citep{meyer1995coordinate}. Note that this run size is two more than that of our 7-factor cOMARS design in Table~\ref{tab:OMARS_design}. Therefore, the simulation results for the D-optimal design reflect both the effect of design type and the effect of two additional runs in the extraction experiment.

Bayesian D-optimal designs \citep{DuMouchel1994} maximize the efficiency of estimating selected (primary) coefficients in the regression model while allowing us to study the other (secondary) coefficients. These designs have a tuning parameter called the prior variance, which sets the trade-off between the design's estimation efficiency for the primary coefficients and its ability to detect the secondary ones. Larger values of the prior variance result in a Bayesian D-optimal design with a greater emphasis on the secondary coefficients. Bayesian D-optimal designs have run sizes that are more flexible than D-optimal design. This is because the run size of the former can be as small as the number of primary coefficients.

For our simulation study, we construct a Bayesian D-optimal design that resembles the cOMARS design in Table~\ref{tab:OMARS_design}. Specifically, our Bayesian D-optimal design has seven factors and 34 runs. Two of these runs are center runs. To generate the other 32 runs, we use JMP v18 with 1,000 iterations for its coordinate-exchange algorithm for Bayesian D-optimal designs. The primary coefficients are the intercept and the linear and quadratic effects. The secondary coefficients are the two-factor interactions. We use a prior variance equal to 1/16, which is the default value in JMP.  

\subsection{Data analysis methods} \label{sec:data_analysis}

We use the best-known data analysis method for each design. For the standard and Bayesian D-optimal designs, we use the Dantzig selector \citep{candes2007dantzig,phoa2009analysis}, which searches for active effects by solving a linear programming problem. The problem involves the minimization of the sum of the absolute values of the estimated coefficients of a model subject to a constraint on the residuals. Essentially, the constraint sets a maximum absolute value of linear combinations of the residuals given by the model matrix. \cite{draguljic2014screening},  \cite{marley2010comparison}, and \cite{Vazquezetal2022} demonstrate the excellent performance of the Dantzig selector to correctly identify active effects when using D-optimal or Bayesian D-optimal designs. In our simulation study, we use the implementation of the Dantzig selector from \cite{Vazquezetal2022}. Specifically, we perform model selection automatically using the corrected Akaike's information criterion \citep[cAIC;][]{claeskens2008model}. To identify active effects, we first fit the model selected using cAIC. Next, we declare an effect as active if the absolute value of its estimated coefficient exceeds 0.5, the smallest size of an active effect in our simulation protocol.   

To analyze the data from the cOMARS design, we use the method of \cite{Hameedetal2023} which involves two steps. In the first step, the method identifies the active linear effects using t-tests and an unbiased estimate of the error's variance obtained by exploiting the orthogonality structure of OMARS designs. In the second step, the method uses forward selection to identify active interactions and quadratic effects. In this step, we can use weak or strong effect heredity \citep{wu2011experiments} to select interactions. That is, we can select an interaction only if one or both of the linear effects of the factors involved were declared active in the first step. A similar restriction can be imposed on the quadratic effects by allowing them to be selected only if the corresponding factor has an active linear effect. The method of \cite{Hameedetal2023} involves various significance levels to identify the active effects. In our simulation study, we use the levels recommended by these authors. Additionally, our version of the method uses effect heredity to find active second-order effects. For interactions, the type of heredity used is weak effect heredity. \revb{In any case, supplementary Section~C shows that the two-step method of \cite{Hameedetal2023} remains competitive also in the absence of heredity constraints.}

\subsection{Simulation protocol} \label{sec:simulation_protocol}

Our simulation protocol involves five active linear effects, four active two-factor interactions, and $q$ active quadratic effects, with $q =$ 1, 2, and 3. We obtained these numbers of active effects from the final models found by \cite{Maestronietal2018} for the pesticides dilufenican and isofenphos. For an $N$-run 7-factor design, \reva{each of our 1,000 simulations consisted of the following steps:}
\begin{enumerate}
    \item We construct the $N \times 35$ full-quadratic model matrix excluding the intercept column. The matrix has seven linear effect columns, 21 two-factor interaction columns, and seven quadratic effect columns. We denote this matrix as $\mathbf{X}$. 
    \item We randomly select five linear effect columns of $\mathbf{X}$ and associated them with the five active linear effects. We randomly select four two-factor interaction columns of $\mathbf{X}$ subject to the constraint that they involved at least one factor with an active linear effect. In other words, we assume that the interactions satisfy weak effect heredity \citep{wu2011experiments}. We randomly select $q$ quadratic columns of $\mathbf{X}$ subject to the constraint that the factor has an active linear effect. 
    \item We generate the coefficient corresponding to an active linear effect by adding 0.5 to an exponentially distributed random number. We generate the coefficient corresponding to an active second-order effect by adding a 1 to an exponentially distributed random number. A ``$+$'' or ``$-$'' sign is randomly assigned to each sampled value. The coefficients of the inactive effects are set to zero.
    \item We generate an $N \times 1$ response vector $\mathbf{y}$ using the model $\mathbf{y} = \mathbf{X} \boldsymbol{\beta} + \boldsymbol{\epsilon}$, where the $35 \times 1$ vector $\boldsymbol{\beta}$ has the simulated coefficients for the active and inactive effects, and the $N \times 1$ vector of errors $\boldsymbol{\epsilon}$ has elements $\epsilon_i$ drawn from $N(0,1)$.
    \item Using $\mathbf{X}$ and $\mathbf{y}$, we identify the active effects using the data analysis method in Section~\ref{sec:data_analysis} that corresponds to the design involved. That is, we use the method of \cite{Hameedetal2023} for the cOMARS design and the Dantzig selector for the other designs. \reva{For each method/design pair, we report the effects declared active and inactive. We also report the least squares estimates of the coefficients for the effects declared as active, since those for the inactive effects are zero.}  
\end{enumerate}

Following the simulation protocols of \cite{Vazquezetal2022} and \cite{Hameedetal2023}, we simulate coefficients for active effects using an exponential distribution. \reva{In this way, our protocol simulates screening scenarios where the signal-to-noise ratio for many active effects is small, while, for a few effects, it is large. Recall that the signal-to-noise ratio is $|\beta_j|/\sigma$, where $\beta_j$ is the simulated coefficient for the $j$-th effect and $\sigma$ is the standard deviation of the error. For an active linear effect, the probability that the signal-to-noise ratio will be between 0.5 and 1.5 is 63.21\%, while the probability that it will be higher than 1.5 is 37.79\%. For an active second-order effect, the probability of 63.21\% applies to a signal-to-noise ratio between 1 and 2, while the probability of 37.79\% applies to a ratio higher than 2. Therefore, our simulation protocol involves hard screening problems where the linear effects are expected to be smaller than the second-order effects. The supplementary material includes R and Python code with the implementation of the protocol.} 

\subsection{Results}

\reva{We use four criteria to compare the designs: power, type-I error, false discovery rate (FDR), and root mean squared error (RMSE) of the least squares estimates of the coefficients. The power is the proportion of active effects that are detected successfully. The type-I error is the proportion of inactive effects that are declared active. The FDR is the proportion of effects declared active that are actually inactive. The RMSE is the square root of the mean squared error of the coefficients' estimates, calculated as the average of the squared differences between the true simulated coefficients and their least squares estimates. Our calculation excludes the intercept and includes the true and estimated coefficients of the active and inactive effects. %%Our calculation for the RMSE includes the simulated coefficients for the inactive effects and the estimates of the effects not selected by the design and analysis strategy, both of which are equal to zero. 

Obviously, the power should be maximized while the type-I error, FDR, and RMSE should be minimized. We compute the four criteria considering all effects jointly and for the linear effects, interactions, and quadratic effects separately. To streamline our discussion, we focus on the power, type-I error, and FDR for the effects separately. Supplementary Section~C shows the simulation results for all effects in terms of the four criteria. It also shows additional simulation results to demonstrate that the method of \cite{Hameedetal2023} is better than the Dantzig selector for analyzing the data from our 7-factor cOMARS design.} 
 
\reva{Table~\ref{tab:simulation_results} shows the average powers, type-I errors, and FDRs of the three designs and their analysis strategies for $1 \le q \le 3$ active quadratic effects. Specifically, the powers, type-I errors, and FDRs are averaged across the 1,000 simulations for each combination of design and number of active quadratic effects.} The table shows that an analysis involving the 32-run cOMARS design has higher average powers for linear effects than an analysis involving the 36-run D-optimal design and the 34-run Bayesian D-optimal design. However, the average powers for two-factor interactions of the non-orthogonal designs are higher than for the cOMARS design. \reva{In any case, the average power for interactions of the cOMARS design is higher than 91\% in the presence of one or two active quadratic effects.} The results in terms of power for linear and interaction effects are in line with the literature on orthogonal and non-orthogonal two-level designs. For instance, \reva{\cite{Meeetal2017} and} \cite{Vazquezetal2022} show that two-level strength-3 orthogonal arrays have higher powers for linear effects but smaller powers for interactions than two-level D-optimal and Bayesian D-optimal designs.

\begin{table}[h]
    \centering
        \caption{Average power, type-I error, and FDR for non-orthogonal designs and cOMARS design with seven factors. $q$: number of quadratic effects}
    \label{tab:simulation_results}
    \begin{tabular}{clccccccccc} \toprule 
    &   &  \multicolumn{3}{c}{Power} & \multicolumn{3}{c}{Type-I error} & \multicolumn{3}{c}{FDR}  \\ \cmidrule{3-11}
$q$	&	Design	&	LE	&	INT	&	QE	&	LE	&	INT	&	QE &	LE	&	INT	&	QE	\\ \midrule
1	&	Bayes D-optimal	&	0.920	&	0.985	&	0.676	&	0.011	&	0.032	&	0.039	&	0.004	&	0.098	&	0.130	\\
	&	D-optimal	&	0.923	&	0.999	&	0.484	&	0.014	&	0.016	&	0.040	&	0.005	&	0.050	&	0.150	\\
	&	cOMARS	&	0.955	&	0.944	&	0.552	&	0.049	&	0.031	&	0.024	&	0.017	&	0.101	&	0.110	\\
2	&	Bayes D-optimal	&	0.904	&	0.976	&	0.644	&	0.010	&	0.035	&	0.037	&	0.003	&	0.110	&	0.081	\\
	&	D-optimal	&	0.921	&	0.999	&	0.426	&	0.015	&	0.020	&	0.028	&	0.005	&	0.063	&	0.071	\\
	&	cOMARS	&	0.964	&	0.912	&	0.510	&	0.049	&	0.043	&	0.024	&	0.016	&	0.142	&	0.075	\\
3	&	Bayes D-optimal	&	0.900	&	0.963	&	0.570	&	0.016	&	0.038	&	0.031	&	0.006	&	0.121	&	0.046	\\
	&	D-optimal	&	0.911	&	0.997	&	0.398	&	0.020	&	0.020	&	0.028	&	0.007	&	0.063	&	0.051	\\
	&	cOMARS	&	0.961	&	0.840	&	0.446	&	0.043	&	0.073	&	0.038	&	0.014	&	0.236	&	0.087	\\
    \bottomrule 
    \end{tabular}
\end{table}

Regarding the power to detect active quadratic effects, \reva{the cOMARS design has a higher power than the D-optimal design for all values of $q$. However, identifying active quadratic effects is challenging for all designs in all cases. For each design, the average power to detect them is less than 0.7 and decreases with the value of $q$.}

\reva{For all designs and numbers of active quadratic effects, the average type-I error rates are smaller than 0.05, which aligns with the standard level used in hypothesis testing in practice. The only exception is the cOMARS design, which has an average type-I error of 0.073 for interactions when $q = 3$. The limited performance of the cOMARS design for this case is also evident from the FDR, since the FDR for interactions is at least twice that of the benchmark designs.} \revb{However, for $q = 1$, the FDRs for interactions of the cOMARS and Bayesian D-optimal designs are similar. Moreover, the FDR for quadratic effects of our design is smaller than the others in this case. For all values of $q$, all FDRs for linear effects in Table~\ref{tab:simulation_results} are less than 2\%.} %%Moreover, for all values of $q$, all FDRs for the linear effects in Table~\ref{tab:simulation_results} are less than 2\%, and the FDRs for quadratic effects of the cOMARS design are strictly smaller than those for the benchmark designs. 534634226

Overall, Table~\ref{tab:simulation_results} shows that the designs do not dominate each other in terms of three of our criteria for linear, interaction, and quadratic effects. Table~C1 in supplementary Section~C further supports this conclusion in terms of our fourth criterion concerning the estimation efficiencies of these effects. In particular, the average RMSEs for the linear effects of the cOMARS design are smaller than for the benchmark designs. However, the \revb{benchmark} designs have a smaller average RMSE for the second-order effects than the \revb{cOMARS design}.

For all designs and their corresponding data analysis strategies, the small powers for quadratic effects suggest that it is challenging to find active quadratic effects with low signal-to-noise ratios. \reva{However, supplementary Section~C shows that setting the minimum signal-to-noise ratio for the active second-order effects to two results in higher average powers than those in Table~\ref{tab:simulation_results}. Specifically, Table~C2 shows that the average powers for the Bayesian D-optimal design and our cOMARS design (both with 34 runs) are higher than 0.74 for the quadratic effects. For the 36-run D-optimal design, the average power for these effects is higher than 0.622. Regarding the interactions, the average powers are higher than 0.94 for all designs in all cases. Table~C2 shows that the average FDRs and RMSEs are generally smaller than those in Table~\ref{tab:simulation_results}. Moreover, all average type-I errors are close to or smaller than 0.05. Based on our simulation results in Tables C1 and C2, we conclude that our 34-run cOMARS design is competitive with the 36-run D-optimal design and the 34-run Bayesian D-optimal design when there are five active linear effects, four active interactions, and one or two quadratic effects.}

\section{Discussion}
\label{sec:discussion}

\revb{In this paper, we constructed new OMARS designs, called cOMARS designs, by concatenating two DSDs. The main merit of the construction method is that it enables researchers to obtain attractive OMARS designs for situations where the OMARS design enumeration algorithm of \cite{NunezAresGoos2020} fails. The method therefore enriches the set of available OMARS design options in the literature. 

We showed} that some statistical properties of the cOMARS designs do not depend on the way the DSDs used are concatenated. Other statistical properties, such as the aliasing between specific second-order effects, do depend on the concatenation of these designs. However, we provided analytical expressions to characterize this aliasing. To minimize the aliasing among the second-order effects, we adapted an existing algorithm developed to concatenate two-level screening designs to deal with three-level designs, and produced good cOMARS designs with seven to 20 factors. We demonstrated that cOMARS designs are better than or competitive with available DSDs and OMARS designs with up to 12 factors. To our knowledge, for 13 to 20 factors, there are no three-level orthogonal designs of comparable sizes as our cOMARS designs in the literature. Therefore, our cOMARS designs provide cost-efficient experimental plans for studying these numbers of factors at three levels, with the property that the first-order effects are not aliased with each other nor with the second-order effects.

Using a simulation study, we demonstrated that a Bayesian D-optimal design, a D-optimal design, and our cOMARS design in Table~\ref{tab:7OMARS} with seven factors did not dominate each other in terms of power, type-I error rate, false discovery rate, and estimation efficiency, measured by the root mean squared error of the coefficients' estimates. Additional simulations (not shown here) revealed that this conclusion holds for problems in which a D-optimal design for studying all first- and second-order effects becomes too expensive due to its run size. Specifically, these simulations involved the 10-factor 41-run cOMARS design labeled ‘f’ in Table~\ref{tab:collection_designs}. As a benchmark, we constructed a Bayesian D-optimal design with 40 runs plus an additional center run using a similar setup as in Section~\ref{sec:nonorthogonal_designs}. The simulation protocol was the same as in Section~\ref{sec:simulation_protocol} except that we used 100 simulations and one and three active quadratic effects for illustrative purposes. The simulation results showed that our 10-factor cOMARS design had a higher average power for linear effects and, for one case, even a higher average power for the interactions than the Bayesian D-optimal design. However, the Bayesian D-optimal design had higher average powers for the quadratic effects than our design.

\revb{Overall, we recommend our cOMARS designs for experiments involving: (i) a budget that does not permit the estimation of the full second-order model; (ii) a moderate or large number of potentially active linear effects; and (iii) a small number of potentially large  active second-order effects. This is mainly due to their run-size economy, well-defined structure, and good aliasing properties of linear effects, and the dedicated data analysis method of \cite{Hameedetal2023}. For experiments with a limited budget and many potentially large and active second-order effects, we recommend three-level Bayesian D-optimal designs analyzed using the Dantzig selector. If the budget allows more runs than the minimum needed to estimate the full second-order model, we recommend three-level D-optimal designs to identify active effects using, for instance, the Dantzig selector.}

We constructed \revb{our} cOMARS designs by concatenating DSDs obtained from conference designs with the smallest run sizes in the catalog of \cite{Schoenetal2022}. However, for a given number of factors, this catalog also includes conference designs with larger run sizes. Using our algorithm, we can therefore also obtain larger cOMARS designs by concatenating DSDs obtained from these larger designs. The analytical formulas and the theoretical result in Section~\ref{sec:properties} also apply to the resulting cOMARS designs. One challenge in our construction of cOMARS designs is to select the best parent DSDs among the available design options. We recommend concatenating the DSDs with as little aliasing between interactions as possible---as quantified using the SSQ or $F$ objective functions in Section~\ref{sec:objective_functions}. Compared to the cOMARS designs constructed here, cOMARS designs constructed using conference designs with larger run sizes will have a smaller aliasing among quadratic and interaction effects but a higher aliasing between quadratic effects. 

%The good statistical properties of our cOMARS designs make them suitable for screening experiments compared to DSDs with similar run sizes. Moreover, the cOMARS designs are competitive with non-orthogonal designs, such as D-optimal designs and Bayesian D-optimal designs, for detecting \reva{large first- and second-order effects.}

\cite{Liuetaletal2022} show that a DSD's D-efficiency to estimate a 3- or 4-factor full second-order response surface model with the first- and second-order effects decreases with its run size. This is due to the limited efficiency of DSDs in estimating the quadratic effects in these models \citep{Vazquezetal2020}. The same is true for cOMARS designs because, for large run sizes, the ratio between the number of zero and non-zero entries will be low in each quadratic effect column. To overcome this issue, we can follow the recommendation of \cite{Liuetaletal2022} and augment cOMARS designs with axial runs, in which all factors are set to their middle level except for a factor that is set to a level outside its pre-established limits. For models with three or four factors, axial runs will reduce the standard errors of the quadratic effects, the bias in the estimated coefficients caused by the omitted terms, and the variance of the response predictions. The decision to augment a cOMARS design with axial runs can be made after analyzing its data using the method of \cite{Hameedetal2023} or the MIO method of \cite{vazquez2021mixed}. If the active effects involve few factors and the goal of the experiment is to build a model that approximates the response surface well, we recommend following up a cOMARS design with axial runs if feasible.

\begin{center}
{\large\bf Supplementary Material}
\end{center}

\begin{itemize}
\item \textbf{Supplementary sections.pdf}. Document with derivations of the properties of cOMARS designs in Section~\ref{sec:properties} and additional properties\reva{, numerical evaluation of the CC/VNS algorithm, and additional simulation results.}
\item \textbf{Programs.zip}. Zip file containing the implementation of the CC/VNS algorithm for optimizing cOMARS designs in Matlab.
\item \textbf{cOMARS designs.zip}. Zip file containing CSV files with cOMARS designs with seven to 20 factors.
\item \textbf{Simulations.zip}. Zip file containing R and Python code to reproduce the simulation study.
\end{itemize}

\begin{center}
{\large\bf Disclosure of Interest}
\end{center}
There are no competing interests associated to this paper.

\begin{center}
{\large\bf Declaration of Funding}
\end{center}
No funding was received.

\begin{center}
{\large\bf Data Availability Statement}
\end{center}
All the data and code concerning this paper is in its supplementary material.

%---------------------------------------------------------------
% Appendix
%---------------------------------------------------------------

%\clearpage

%%\renewcommand{\thesection}{}
%%\section*{Appendix}
%%
%%\appendix
%%\section{Objective function}
%%\label{app:objective}

%---------------------------------------------------------------
% References
%---------------------------------------------------------------
%\clearpage
\bibliographystyle{apalike} % Plain referencing style
\bibliography{referencesOMARS} % Use the example bibliography file sample.bib

\end{document}